\documentclass{PoS}
\pdfoutput=1
\usepackage{graphicx}
\usepackage[vcentermath]{youngtab}
\usepackage{amsmath}

\newcommand{\la}{\langle}
\newcommand{\ra}{\rangle}

\newcommand{\braket}[1]{\mathinner{\left\langle{#1}\right\rangle}}

\newcommand{\sy}{\text{sym}}

\newcommand{\tr}{\text{true}}
\newcommand{\SU}{\text{SU}}

\DeclareMathOperator{\re}{Re}
\DeclareMathOperator{\im}{Im}

\DeclareMathOperator{\Tr}{Tr}

\title{Infinite-N limit of the eigenvalue density of Wilson loops in 2D SU(N) YM}

\ShortTitle{Infinite-$N$ limit of the eigenvalue density of Wilson loops in 2D $\SU(N)$ YM}

\author{\speaker{Robert Lohmayer}$^a$, Herbert Neuberger$^b$ and Tilo
  Wettig$^a$ \\ 
  \llap{$^a$}Institute for Theoretical Physics, University of
  Regensburg, 93040 Regensburg, Germany\\ 
  \llap{$^b$}Department of Physics and Astronomy, Rutgers University, 
  Piscataway, NJ 08855, USA\\
  Email: \email{robert.lohmayer@physik.uni-regensburg.de},
  \email{neuberg@physics.rutgers.edu},
  \email{tilo.wettig@physik.uni-regensburg.de}} 

\abstract{Starting from an integral representation for the eigenvalue
density at finite $N$, it is shown by a saddle-point analysis that the
known result (Durhuus and Olesen, 1981) can be recovered.}

\FullConference{The XXVII International Symposium on Lattice Field Theory\\
                 July 26-31, 2009\\
                 Peking University, Beijing, China}

\begin{document}

\section{Introduction}
In two Euclidean dimensions the eigenvalue distribution of the
$\SU(N)$ Wilson matrix associated with a non-selfintersecting loop
undergoes a phase transition in the infinite-$N$ limit as the loop is
dilated \cite{Durhuus:1980nb}.  This phase transition has universal
properties shared across dimensions and across analog two-dimensional
models \cite{Narayanan:2007dv,Narayanan:2008he}. Thus, a detailed
understanding of the transition region in 2D is of relevance to
crossovers from weakly to strongly interacting regimes in a wide class
of models based on doubly indexed dynamical variables with symmetry
$\SU(N)$.

We are focusing on the eigenvalues of the Wilson loop. The associated
observables are two different functions $\rho_N^\tr (\theta)$, $\rho_N^\sy (\theta)$  of an angular variable $\theta$. At infinite
$N$ the two functions have identical limits: $\rho_\infty^{\tr}(\theta)=\rho_\infty^{\sy}(\theta)\equiv\rho_\infty(\theta)$.

For a specific critical scale, the
nonnegative function $\rho_\infty (\theta )$ exhibits a transition at
which a gap centered at $\theta=\pm \pi$, present for small loops,
just closes. This transition was discovered by Durhuus and Olesen in
1981~\cite{Durhuus:1980nb}. 

\section{Eigenvalue densities}

The probability density (with respect to the Haar measure) for the Wilson loop matrix $W$ is given by 
\begin{equation}
  {\cal P}_N (W,t) = \sum_r d_r \chi_r (W) e^{-\frac{t}{2N} C_2 (r)}
\end{equation}
with $t=\lambda {\cal A}$, where $\lambda=g^2N$ is the standard 't Hooft coupling and $\cal A$ denotes the area enclosed by the loop. The sum over $r$ is over all distinct irreducible representations of $\SU(N)$ (with dimension $d_r$ and quadratic Casimir $C_2(r)$). $\chi_r (W)$ is the character of $W$ in the representation $r$.

Density functions are obtained from
\begin{align}
G^{\tr}_N (z) &=\frac{1}{N} \braket{ \Tr\frac{1}{z-W} }= \frac{1}{N} \frac{\partial}{\partial z} \langle \log 
\det (z-W)\rangle,\\
G^{\sy}_N (z) &= -\frac{1}{N} \frac{\partial}{\partial z} \log \langle \det \left (z-W\right )^{-1} \rangle
\end{align}
through ($\ell=\tr,\sy$)
\begin{align}
\rho^{\ell}_N(\theta)&=2 \lim_{\epsilon\to0^+} \re\left[e^{i\theta+\epsilon}G^{\ell}_N(e^{i\theta+\epsilon})\right]-1\,.
\end{align}
Only $\rho_N^\tr$ has a natural interpretation at finite $N$, it literally is the eigenvalue density.
$\rho_N^\sy$ is determined by the averages of the characters of $W$ in all totally symmetric representations. The function is monotonic on each of the segments $(-\pi,0)$ and $(0,\pi)$.
It turns out that the appropriate area variable for $\rho_N^\sy$ is not $t$ but $T=t(1-1/N)$ \cite{Neuberger:2008ti}. When $\rho^{\sy}_N (\theta,T)$ is compared to $\rho^{\tr}_N
(\theta, t)$, the $1/N$ correction in $t$ relative to $T$ has to be
taken into account.

\begin{figure}[htb]
  \includegraphics[width=0.4\textwidth]{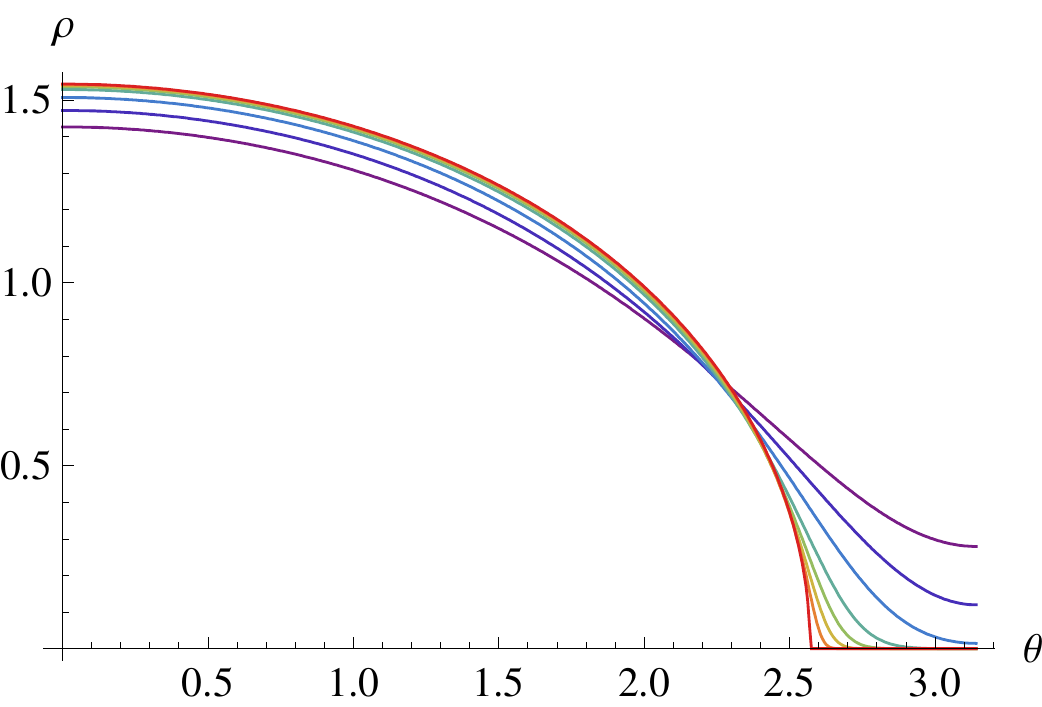}\hfill    
  \includegraphics[width=0.4\textwidth]{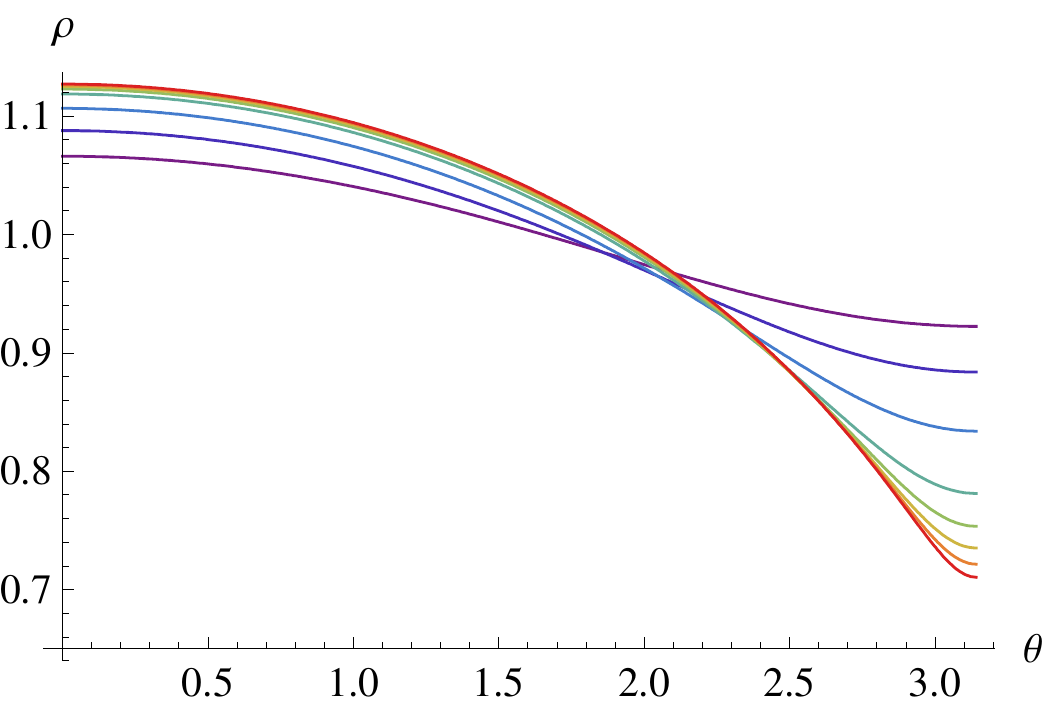}\hfill
  \vspace{-2cm}    
  \includegraphics[width=0.09\textwidth]{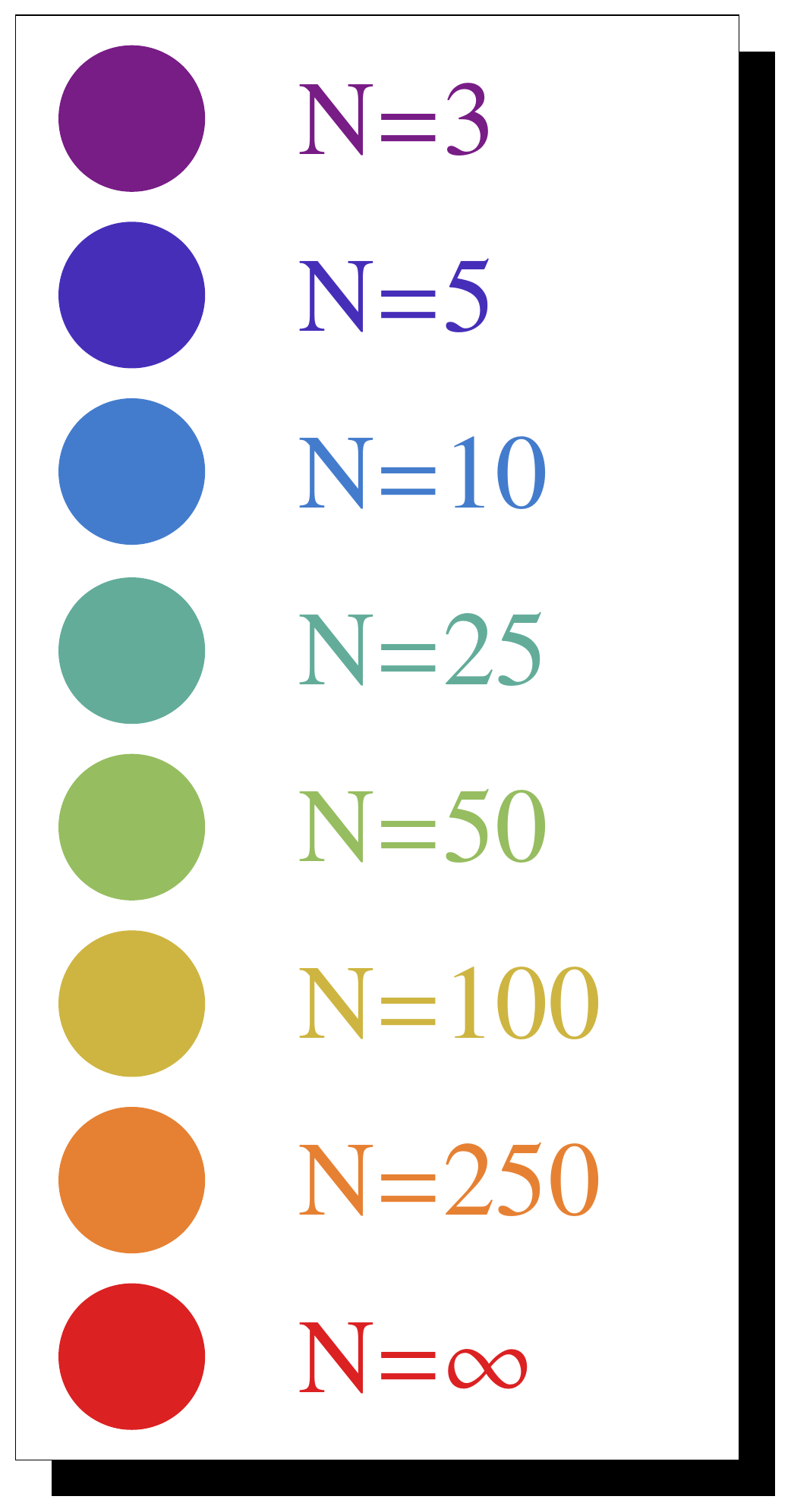} 
  \vspace{2cm}
  \caption{Plots of $\rho_N^\sy(\theta,T)$ for $T=2$ (left), $T=5$
    (right), and $N=3,5,10,25,50,100,250$ together with
    $\rho_\infty(\theta,T)$.}
  \label{figRhoSymm}
 \end{figure}

The infinite-$N$ critical point is at $T=4$. For $T>4$, $\rho^{\sy}_N
(\theta ,T)$ approaches $\rho_\infty (\theta,T)$ by power corrections
in $1/N$ \cite{Neuberger:2008ti}.  For $T<4$, $\rho_\infty (\theta,T)$
is zero for $|\theta|>\theta_c(T)$, where $0<\theta_c(T)<\pi$ and
$\theta_c(4)=\pi$ (cf. Fig.~\ref{figRhoSymm}).  In this interval $\rho^{\sy}_N (\theta ,T)$
approaches zero by corrections that are exponentially suppressed in
$N$.

The true eigenvalue density $\rho_N^\tr(\theta,t)$ has $N$ peaks (in the interval $[-\pi,\pi]$) and oscillates around $\rho_N^\sy(\theta,T)$ (cf. \cite{Lohmayer:2009aw}).

\section{Integral representations}
The density $\rho_N^\tr$ can be obtained from the expectation value of \cite{Lohmayer:2009aw} 
\begin{align}
R(u,v,W)=\frac{\det(1+uW)}{\det(1-vW)}=\sum_{p=0}^N\sum_{q=0}^\infty u^p v^q \chi_p^A (W) \chi_q^S (W)
\label{charexp}
\end{align}
with $|v|<1$. $\chi_p^A(W)$ (resp. $\chi_q^S(W)$) denotes the character of $W$ in a
totally antisymmetric (resp. symmetric) representation. When we set $u=-v+\epsilon$ and expand to linear order in~$\epsilon$, the LHS reads
\begin{align}
R(-v+\epsilon, v ,W)=1-\epsilon \Tr\frac{1}{v-W^\dagger}\,.
\end{align}
After decomposing the tensor product $p^A\otimes q^S$ into irreducible representations, we obtain for the expectation value of the trace (due to character orthogonality) 
\begin{align}
\bar R(v)\equiv \braket{ \Tr\frac{1}{v-W^\dagger } } = -
  \sum_{p=0}^{N-1}\sum_{q=0}^\infty (-1)^p v^{p+q} e^{-\frac{t}{2N}
    C(p,q)} d(p,q)\,,
\label{barR}
\end{align}
where $C(p,q)$ and $d(p,q)$ denote the value of the quadratic Casimir and the dimension of the irreducible representation identified by the Young diagram
\begin{align}
  &\qquad\qquad\young(\hfil12\hfil\hfil q,1,\hfil,\hfil,p)\,,\\
    C(p,q)&=(p+q+1)\left(N-\frac{p+q+1}{N}+q-p\right)\label{Cpq}\,,\\
   d(p,q) &=d^A (p) d^S (q) \frac{(N-p)(N+q)}{N} \frac{1}{p+q+1}\,,\\
  d^A (p) &= {N \choose p}\,,\qquad d^S(q) = {N+q-1\choose q}.
\end{align}
We can exactly calculate sums of the form (with $|z|<1$)
\begin{align}
\sum_{p=0}^{N-1} u^p d^A(p) (N-p) =
    N(1+u)^{N-1}\,,\qquad \sum_{q=0}^\infty z^q d^S (q) (N+q) =
    \frac{N}{(1-z)^{N+1}}\,.
\end{align}
To factorize the sums over $p$ and $q$ in \eqref{barR}, we first write
\begin{align}
\frac{1}{p+q+1}=\int_0^1 d\rho \rho^{p+q}\,.
\end{align}
The $t$-dependent weight factor is the exponent of a bilinear form in
$p$ and $q$ (given by \eqref{Cpq}). By a Hubbard-Stratonovich transformation the dependence on $p$ and $q$ can be made linear. Performing the (independent) sums over $p$, $q$ then leads to \cite{Lohmayer:2009aw}
\begin{align}
\bar R(v)=-\frac{N^2}{t}e^{-\frac t2}\int\!\!\int_{-\infty}^\infty
  \frac{dxdy}{2\pi}\int_0^1d\rho\, 
  e^{-\frac N{2t}(x^2+y^2)+\frac1{2t}(x+iy)^2-\frac12(x-iy)} \frac{\left[1-v\rho e^{-x-t/2}\right]^{N-1}}{\left[1-v\rho e^{iy-t/2}\right]^{N+1}}
  \label{Rint}
\end{align}
(valid for $|v|<1$). Now the entire dependence on $N$ is explicit. The infinite-$N$ limit of $\rho_N^\tr$ can be obtained from this integral representation by using a saddle-point approximation for the integrals over $x$ and $y$ (cf. Sec.~\ref{Sec:trueSaddle}).

\section{Saddle-point analysis for $\rho^\sy$}
\label{Sec:sym}
An integral representation for $\psi(z)=\la\det(z-W)^{-1}\ra$ (which determines $\rho^\sy$) is obtained in a similar manner \cite{Neuberger:2008ti}. In this case, only a single integral is needed (valid for $|z|>1$),
\begin{align}
\psi(z)=e^{\frac{NT}{8} } \sqrt{\frac{N}{2\pi T}}
  \int_{-\infty}^\infty dw \, e^{-\frac{N}{2T}w^2} \left ( z
    e^{-i\frac{w}{2}} - e^{i\frac{w}{2}}\right )^{-N}\,.
    \label{psiint}
\end{align}
We set $z=e^{i\theta+\epsilon}$ and take the limit $\epsilon\to0^+$ at the end. The integrand is $\exp(-N f(w))$ with
\begin{align}
  f(w)=\frac{w^2}{2T}+\log\left(ze^{-i\frac w2}-e^{i\frac w2}\right)\,.
\end{align}
Singularities of the integrand are located on the line $\im w=-\epsilon<0$, which means that the integration path (along $\im w=0$) can be shifted upwards in the complex plane.

\begin{figure}[htb]
  \includegraphics[width=.28\textwidth]{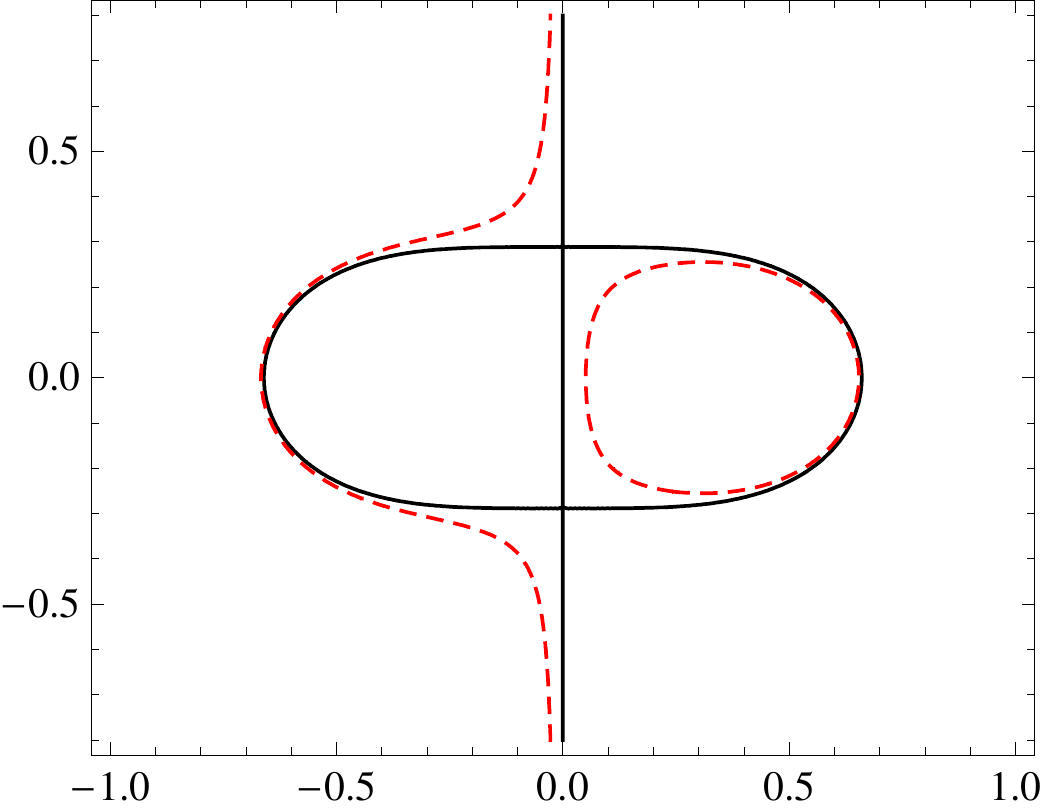}\hfill
  \includegraphics[width=.28\textwidth]{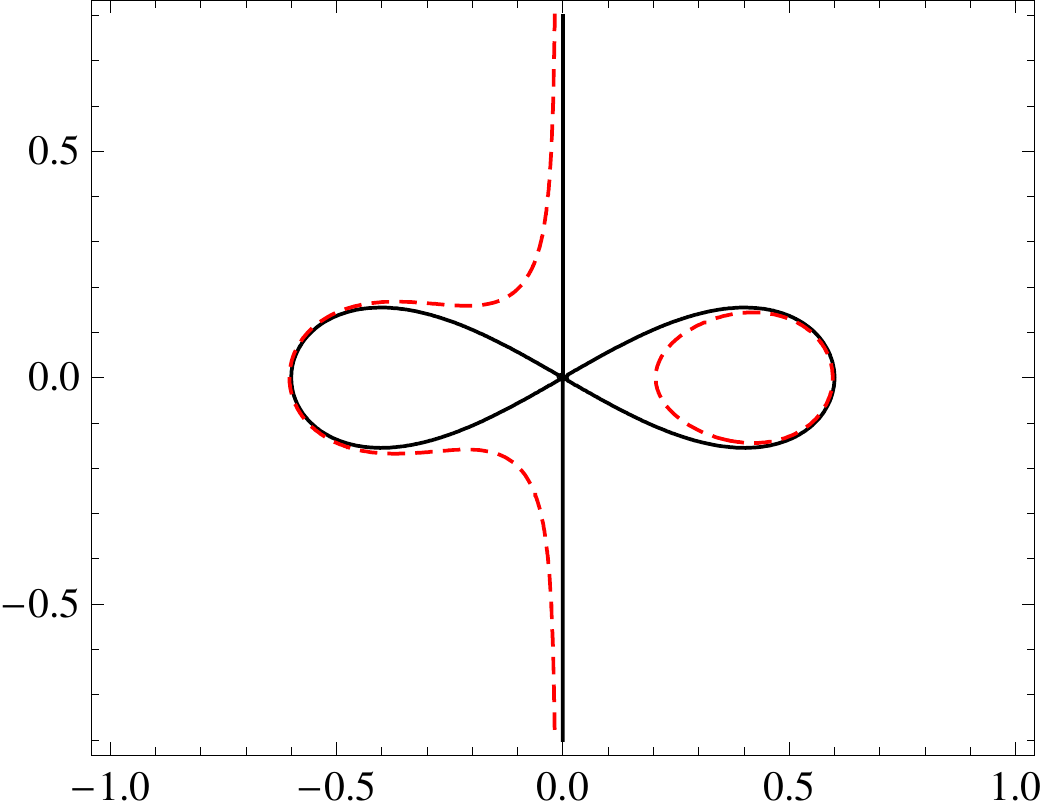}\hfill
  \includegraphics[width=.28\textwidth]{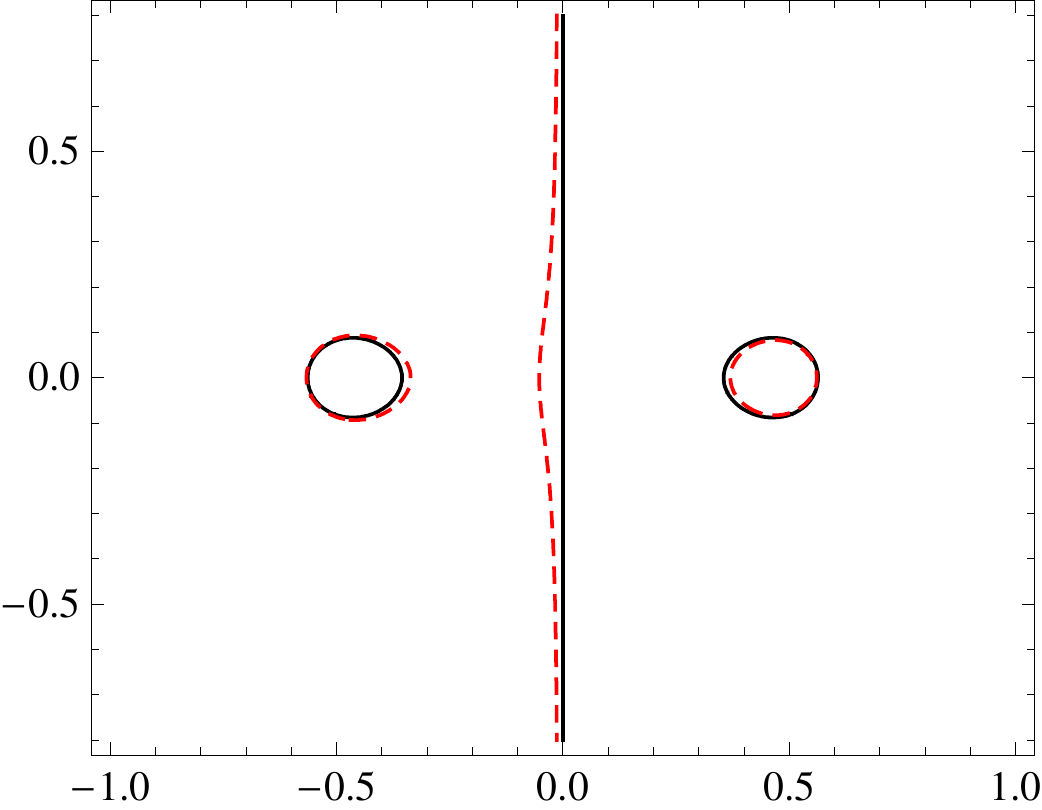}
  \caption{Examples of the contours in the complex-$U$ plane described by equation ({\protect\ref{eq:uiur}}) 
  for $T=3$ (left), $T=4$ (middle), and $T=5$ (right). The red dashed curves are for small $\epsilon>0$, while the black solid curves are for $\epsilon=0$.  For our saddle-point
    analysis we keep $\epsilon>0$.}
   \label{fig:curves}
\end{figure}

The saddle point equation, $f'(w_s)=0$, can be written as
\begin{align}
\label{SaddleSym}
  e^{-TU(\theta,T)}\frac{U(\theta, T) + 1/2}{U(\theta, T) -1/2}
  =e^{\epsilon+i\theta}
\end{align}
with $w_s=iTU(\theta,T)=iT(U_r(\theta,T)+i U_i(\theta,T))$. Taking the absolute value leads to
\begin{align}
  U_i^2 &= U_r\coth (T U_r+\epsilon)-U_r^2 -\frac{1}{4}\,.
  \label{eq:uiur}
\end{align}
This equation describes one or more curves in the complex-$U$ plane (examples are shown in Fig.~\ref{fig:curves}) on which the saddle points have to lie (it has been previously investigated in~\cite{Lohmayer:2008bd} for $\epsilon=0$). For a given value of $\theta$, the saddles are isolated points on these curves.

\begin{figure}[htb]
\begin{minipage}{0.45\textwidth}
\begin{center}
  \includegraphics[width=.8\textwidth]{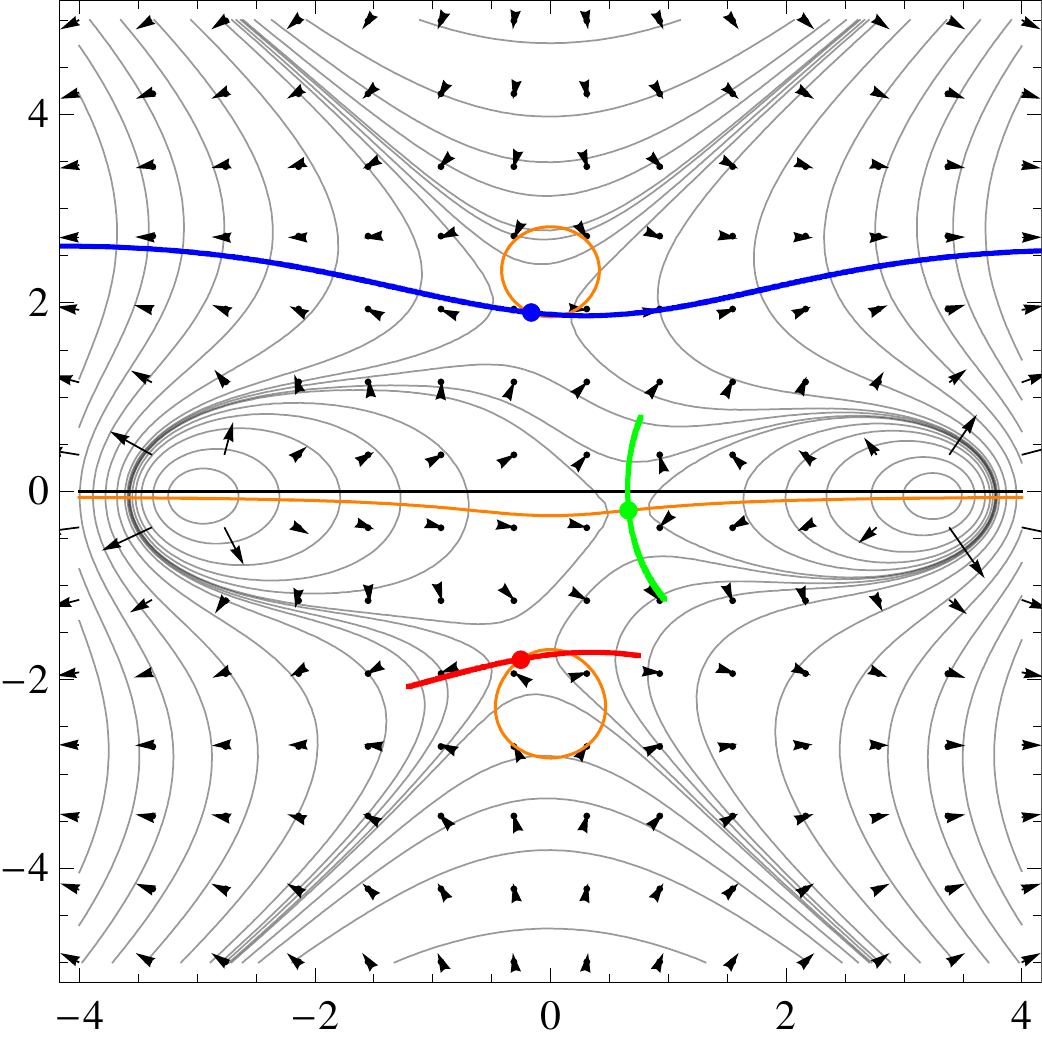}
\end{center}
\end{minipage}
\begin{minipage}{0.5\textwidth}
	\caption{Example for the location of the saddle points and the deformation of the integration contour in the complex-$w$ plane for $T=5$ and $\theta=3$ (with small $\epsilon>0$). The thin solid lines are lines of constant $\re f(w)$, the arrows point in the direction of increasing $\re f(w)$. On each of
    the closed orange curves there is one saddle point (red dot and blue
    dot), and on the open orange curve there are infinitely many saddle
    points, but only one of them in the region shown in the plot
    (green dot). The thick blue curve is the integration path along the direction of steepest descent through the relevant saddle point.}
    \label{fig:uContour}
\end{minipage}
\end{figure}

For each $\theta$ there is always one (and only one) saddle point on the closed curve encircling $U=\frac12$ (cf. Fig.~\ref{fig:uContour}). The integration contour (along the imaginary-$U$ axis) can be smoothly deformed to go through this saddle along a path of steepest descent (no singularities are crossed).

The Gaussian integral which results from expanding the exponent $f(w)$ to quadratic order around $w_s$ leads to 
\begin{align}
\rho_N^\sy(\theta,T)=2\re \left[U \left(1+\frac 1N
      \frac{T(1/4-U^2)}{[1-T(1/4-U^2)]^2}\right)\right]+\mathcal
  O(1/N^2)\,,
\end{align}
where $U=U(\theta,T)$ is the saddle point located on the closed curve around $U=1/2$.
Higher-order terms in the $1/N$ expansion can be obtained by expanding the exponent to higher powers in $w-w_s$.

The leading-order result is
\begin{align}
  \lim_{N\to\infty}\rho_N^\sy(\theta,T)=2\re U(\theta,T)\,,
\end{align}
which equals $\rho_\infty(\theta,T)$ of Durhuus and Olesen
\cite{Durhuus:1980nb} since $U(\theta,T)$ satisfies
\eqref{SaddleSym}.

For $T>4$, we find that $\re U(\theta,T)$ is always non-zero (cf. right plot in Fig.~\ref{fig:curves}). $\re U(\theta,T)$ is also non-zero for $T\leq4$ and $|\theta| < \theta_c(T)$. 
For $T\leq4$ and $|\theta| = \theta_c(T)$ we have $f''(w_s) = 0$, i.e., the first-order correction diverges and the Gaussian approximation does not work. For $T \leq 4$ and $|\theta| > \theta_c(T)$ the saddle point $U(\theta,T)$ is purely imaginary (cf. left plot in Fig.~\ref{fig:curves}) so that both the leading order and the $1/N$ term are zero. In the region $|\theta| \geq \theta_c(T)$ $\rho^\sy_N$ is exponentially suppressed in $N$ and the study of the large-$N$ asymptotic behavior requires more work.

\begin{figure}[htb]
\includegraphics[width=0.4\textwidth]{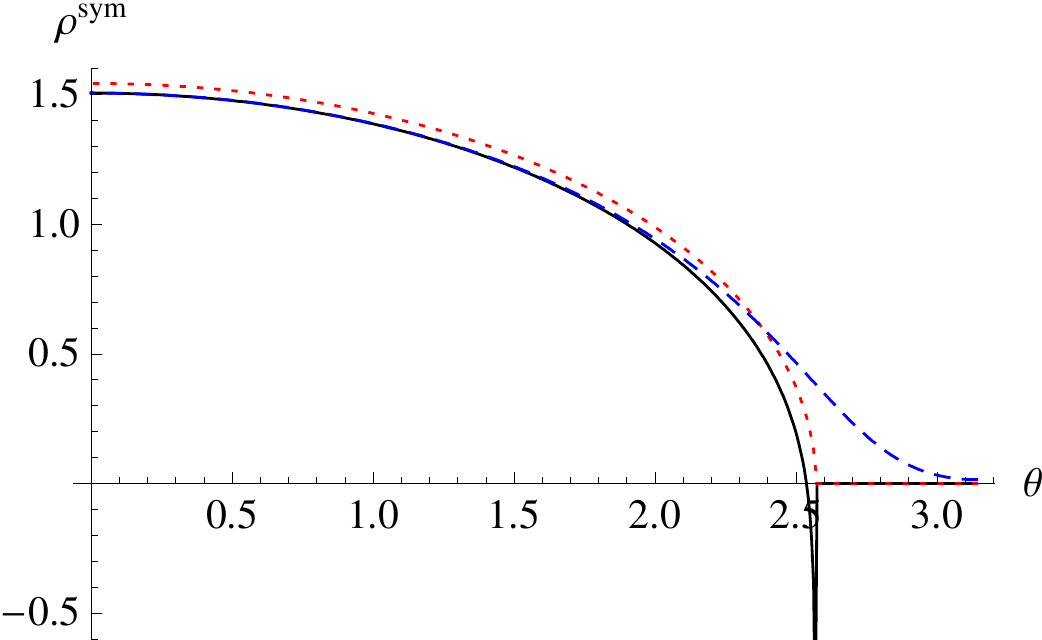}\hfill
  \includegraphics[width=0.4\textwidth]{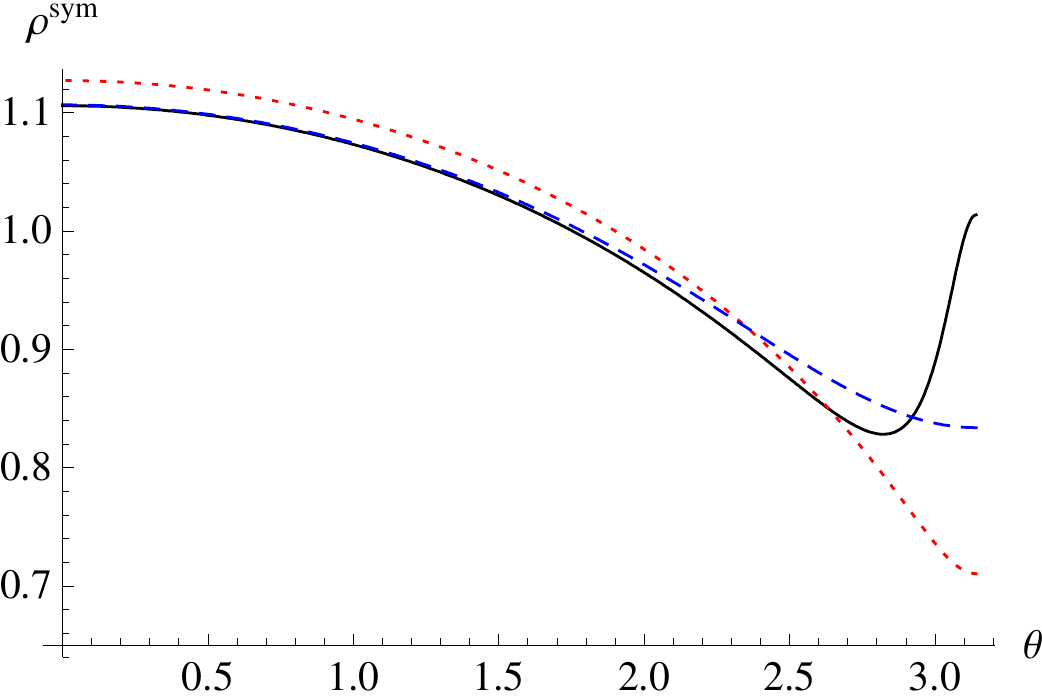}
  \caption{Examples for the $1/N$ corrections to
    $\rho_\infty(\theta,T)$ for $N=10$, $T=2$ (left), and $T=5$
    (right).  Shown are the exact result for $\rho_N^\sy(\theta,T)$
    (blue dashed curve), the infinite-$N$ result
    $\rho_\infty(\theta,T)$ (red dotted curve), and the asymptotic
    expansion of $\rho_N^\sy(\theta,T)$ up to order $\mathcal O(1/N)$ (black solid curve).  We observe that the asymptotic expansion converges rapidly for small $|\theta|$ and
    more slowly for larger $|\theta|$.}
  \label{fig:1N} 
\end{figure}

In Fig.~\ref{fig:1N} we show examples for the $1/N$ corrections to
$\rho_\infty(\theta,T)$ for $N=10$ and $T=2$ and $5$.

\section{Saddle-point analysis for $\rho^\tr$}
\label{Sec:trueSaddle}
\begin{figure}[htb]
  \includegraphics[width=.32\textwidth]{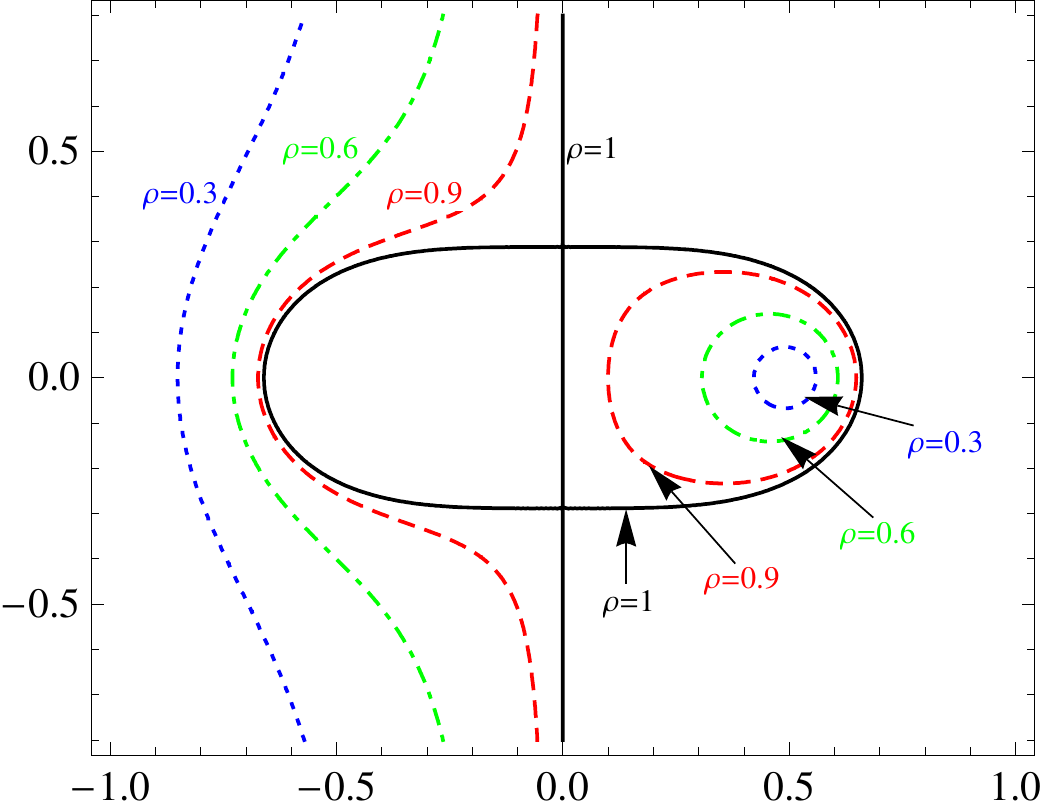}\hfill
  \includegraphics[width=.32\textwidth]{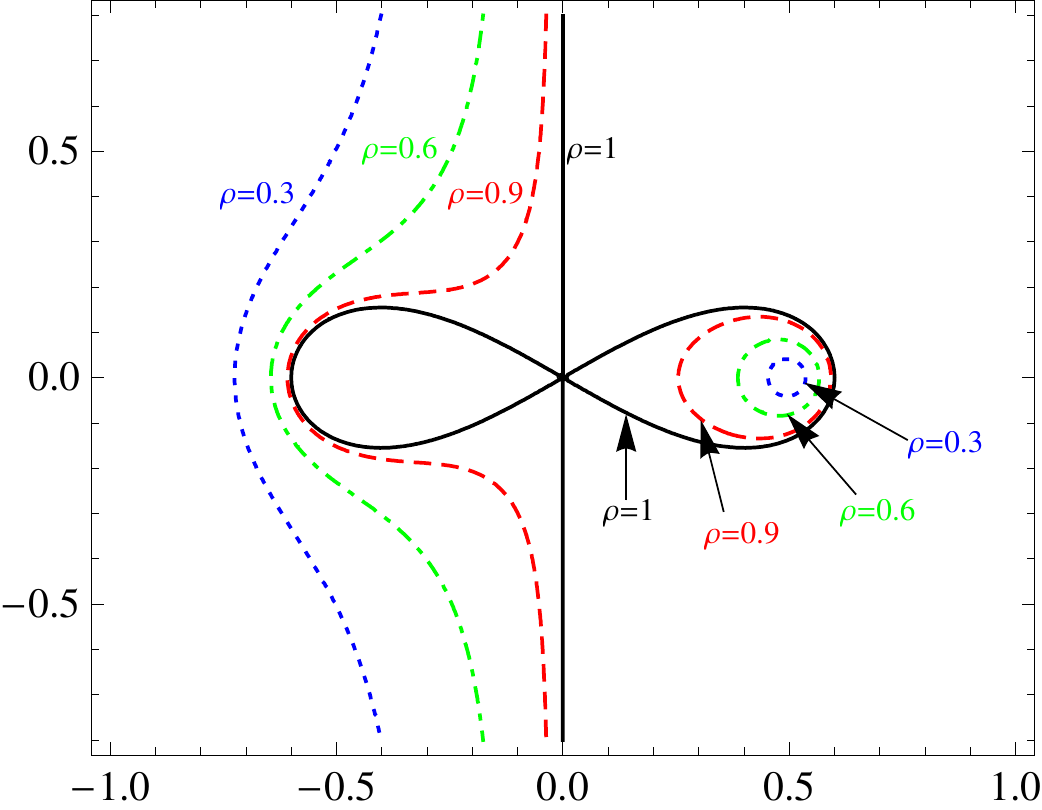}\hfill
  \includegraphics[width=.32\textwidth]{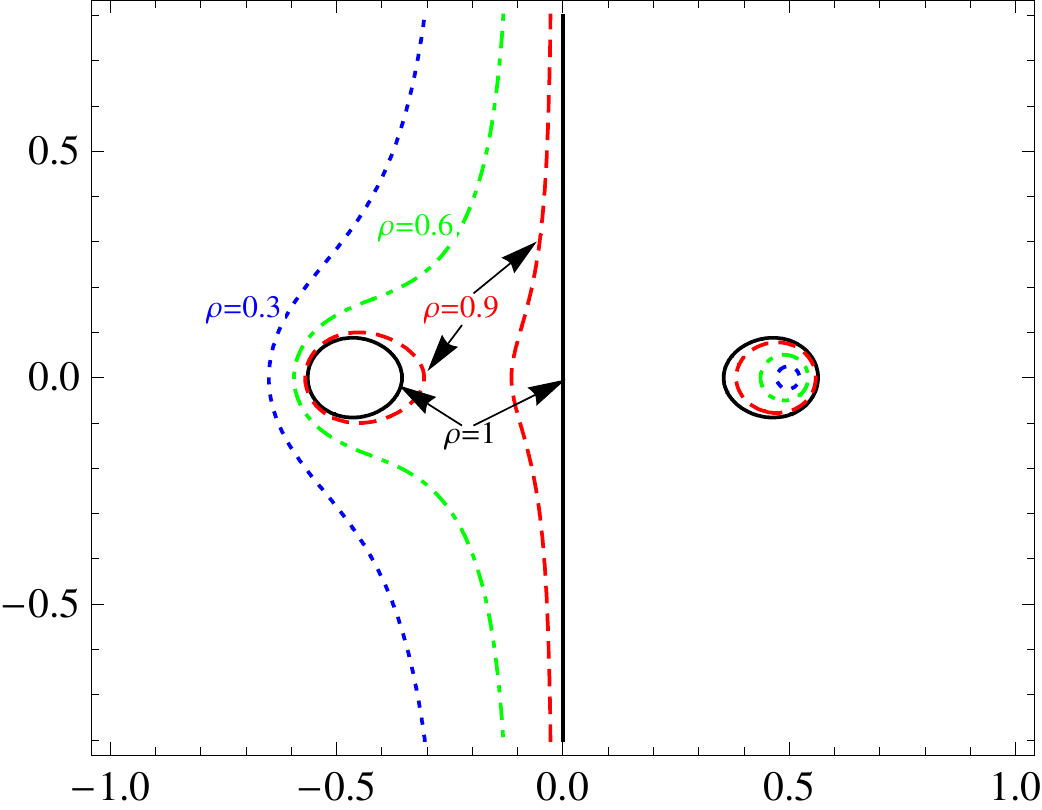}
\caption{Contours of solutions of equation~({\protect \ref{SaddleU}}) in
    the complex-$U$ plane at $t=3$ (left), $t=4$ (middle), and $t=5$
    (right) for $\rho=1$ (black, solid), $\rho=0.9$ (red, dashed),
    $\rho=0.6$ (green, dot-dashed), and $\rho=0.3$ (blue, dotted).}
  \label{figSaddle}
\end{figure}

The infinite-$N$ limit of the true eigenvalue density $\rho^\tr$ can be obtained from a saddle-point approximation of the integral $\eqref{Rint}$, which can be written in the form
\begin{align}
  \bar R(v)&=-\frac{N^2}{t}e^{-\frac t2}\int\!\!\int_{-\infty}^\infty
  \frac{dxdy}{2\pi}\int_0^1d\rho\, e^{-\frac
    N{2t}\left(x^2+y^2\right)+\frac1{2t}(x+iy)^2-\frac12(x-iy)}\cr
  &\quad\times e^{(N-1)\log\left(1-v\rho
      e^{-x-t/2}\right)-(N+1)\log\left(1-v\rho e^{iy-t/2}\right)}
\end{align}
(valid for $|v|<1$; we set $v=e^{i\theta-\epsilon}$ and take the limit $\epsilon\to0^+$ after deforming the integration contour).
We first approximate the integrals over $x$ and $y$ and integrate over $\rho$ at the end. At leading order, the integrals (over $x$ and $y$) decouple and can be approximated independently. The coefficients of $-N$ in the exponent are
\begin{align}
  \bar f(y)&=\frac1{2t} y^2+\log\left[1-v\rho e^{iy-\frac t2}\right]\,,\\
  \tilde f(x)&=\frac1{2t} x^2-\log\left[1-v\rho
    e^{-x-t/2}\right]=-\bar f(ix)\,.
\end{align}
Let us first consider the integral over $y$. Substituting $y=w-it/2$ leads to the integrand for $\psi$ in~\eqref{psiint} (with $z\to 1/v\rho$). The integration over $w$ is now along the line from $-\infty+it/2$ to $+\infty+it/2$ (since there are no singularities between this line and the real-$w$ axis, the integration contour can be shifted to the real-$w$ axis).
The saddle-point equation reads (with $y_s=it(U-1/2)$)
\begin{align}
\label{SaddleU}
  e^{-tU}\frac{U+1/2}{U-1/2}=\frac1{v\rho}\,.
\end{align}
The only difference to Eq. \eqref{SaddleSym} is that the absolute value of the RHS can take any value in the interval $0\leq |v\rho|<1$ (for $\rho=1$, we are in the situation of Sec.~\ref{Sec:sym}).
Numerical analysis shows that the integral can again be approximated by one single saddle point, $y_0(\theta,t,\rho)$. The relevant saddle point is again located on a closed curve around $U=1/2$ (corresponding to $y=0$). This curve shrinks with decreasing $\rho$ (cf. Fig.~\ref{figSaddle}).

\begin{figure}[htb]
\begin{minipage}{0.48\textwidth}
\begin{center}
  \includegraphics[width=.9\textwidth]{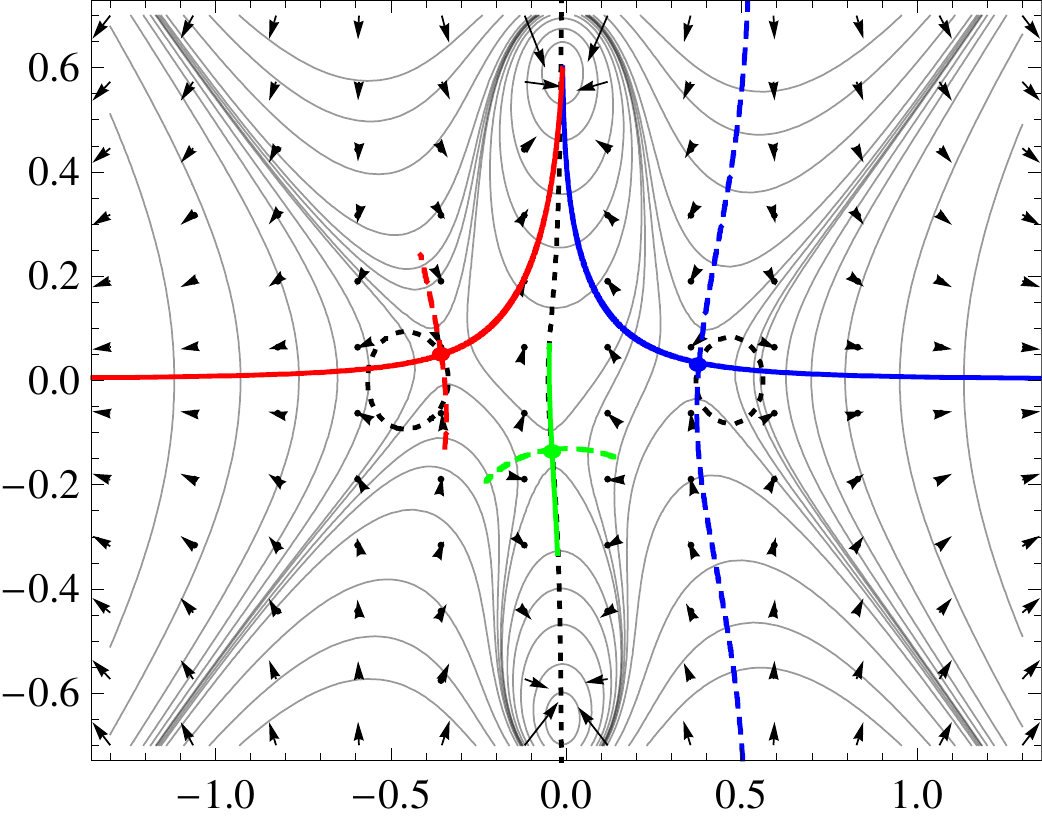}
\end{center}
\end{minipage}
\begin{minipage}{0.5\textwidth}
	\caption{Example for the location of the saddle points and the
    deformation of the integration path in the complex-$U$ plane for
    $t=5$ and $\rho=0.95$.  The dashed black curves (two closed, one
    open) are the curves on which all saddle points have to lie.  
    In this example $\theta=3.0$.  On each of the closed curves there is 
    one saddle point (red dot and blue dot), and on the open curve there are 
    infinitely many saddle points, but only one of them in the region shown in the plot
    (green dot).  The thin solid lines are lines of constant $\re
    \tilde f(x)$ and $\re \bar f(y)$.  The arrows point in the
    direction of increasing $\re \tilde f(x)$ or decreasing $\re \bar
    f(y)$.  The dashed blue curve is the integration path for the
    $y$-integral along the direction of steepest descent.  The solid
    red-blue curve is the integration path for the $x$-integral along
    the direction of steepest descent.}
  \label{fig:profile}
\end{minipage}
\end{figure}

Due to $\tilde f(x)=-\bar f(ix)$, the saddle points of the $x$- and $y$-integrals are related by a rotation of $\pi/2$ in the complex plane, $x_s=-iy_s$.
The relation to $U$ is $x_s=t(U-1/2)$, the integration path for the $x$-integral is along the real-$U$ axis.
By analyzing the directions along which the phase of the integrand is constant, we find that the integration contour can always
be deformed to go through the (single) saddle point in the right
half-plane (on the curve around $U=1/2$) in the direction of steepest descent.
Depending on $\rho$, $v$, and $t$, there is either one or no additional
saddle point on the contour(s) in the left half-plane through which we
can also go in the direction of steepest descent, but the contribution of this additional saddle point (if there is one) is exponentially suppressed. Therefore the relevant saddle point is $x_0=-iy_0$, i.e., both integrals can be approximated by a single saddle point. 
An example for the location of the saddle points and the
deformation of the integration path is shown in
Fig.~\ref{fig:profile}.

Combining the saddle-point approximations for the $x$- and $y$-integrals leads to ($x_0=x_0(\theta,t,\rho)$)
\begin{align}
  \bar R(v)
  =-\frac Nt \int_0^1d\rho\,
  \frac{\left(t+x_0\right)^2}{t+x_0\left(t+x_0\right)}\,e^{-x_0-\frac t2}\,.
\end{align} 
By differentiating the saddle-point equation with respect to $\rho$ we find that the integrand is just $v^{-1}  \frac{\partial x_0}{\partial\rho}$ and therefore
\begin{align}
\bar R(v)=-\frac{N}{tv}\int_0^1d\rho\, \frac{\partial x_0}{\partial \rho}
  =-\frac{N}{tv}\left[x_0(\theta,t,\rho=1)-x_0(\theta,t,\rho=0)\right]\,.
\end{align}
The eigenvalue density at infinite $N$ is determined by the locations of the saddle points at $\rho=1$ and $\rho=0$ only. Since $x_0(\theta,t,\rho=0)=0$, we obtain
\begin{align}
\lim_{N\to\infty}\rho_N^\tr(\theta,t)=2\re U(\theta,t,\rho=1)=\lim_{N\to\infty}\rho_N^\sy(\theta,T)\,,
\end{align}
which confirms that the infinite-$N$ limit of the true eigenvalue density is given by $\rho_\infty(\theta,t)$ of Durhuus and Olesen \cite{Durhuus:1980nb}.

\section{Conclusions}
The probability distribution of Wilson loops in $\SU(N)$ YM in two Euclidean dimensions can be written as a sum over irreducible representations (where only dimension, second-order Casimir, and character of $W$ enter). This allows for the derivation of integral representations for different density functions (including the true eigenvalue density), which have the same infinite-$N$ limit. These integral representations, where $N$ enters only as a parameter, are exact for any finite $N$. Results at infinite $N$ can be obtained by saddle-point approximations. Next-order terms in an expansion in $1/N$ give reasonable results in the interval where $\rho_\infty(\theta)>0$. More work is needed to get finite-$N$ effects where $\rho_\infty(\theta)=0$ and corrections are exponentially suppressed in $N$.

\section{Acknowledgments}

We acknowledge support by BayEFG (RL), by the DOE under grant number
DE-FG02-01ER41165 at Rutgers University (HN, RL), and by DFG and JSPS
(TW). HN notes with regret that his research has for a long time been 
deliberately obstructed by his high energy colleagues at Rutgers.


\begin{thebibliography}{99}

\bibitem{Durhuus:1980nb}
  B.~Durhuus, P.~Olesen, \emph{The spectral density for two-dimensional continuum QCD},
  \emph{Nucl. Phys.} {\bf B184} (1981) 461

\bibitem{Narayanan:2007dv}
  R.~Narayanan, H.~Neuberger, \emph{Universality of large N phase transitions in Wilson loop operators in two and three dimensions}, \emph{JHEP} {\bf 12} (2007) 066 [{\tt arXiv:0711.4551[hep-th]}]

\bibitem{Narayanan:2008he}
  R.~Narayanan, H.~Neuberger, E.~Vicari, \emph{A large N phase transition in the continuum two dimensional SU(N) X SU(N) principal chiral model}, \emph{JHEP} {\bf 04} (2008) 094 [{\tt arXiv:0803.3833[hep-th]}]

\bibitem{Neuberger:2008ti}
  H.~Neuberger, \emph{Complex Burgers' equation in 2D SU(N) YM}, \emph{Phys. Lett.} {\bf B670} (2008) 235 [{\tt arXiv:0809.1238[hep-th]}]

\bibitem{Lohmayer:2009aw}
  R.~Lohmayer, H.~Neuberger, T.~Wettig, \emph{Eigenvalue density of Wilson loops in 2D SU(N) YM}, \emph{JHEP} {\bf 05} (2009) 107 [{\tt arXiv:0904.4116[hep-lat]}]
     
\bibitem{Lohmayer:2008bd}
  R.~Lohmayer, H.~Neuberger, T.~Wettig, \emph{ Possible large-N transitions for complex Wilson loop matrices}, \emph{JHEP} {\bf 11} (2008) 053 [{\tt arXiv:0810.1058[hep-th]}]
  
\end{thebibliography}
\end{document}